\documentstyle[PASJadd]{PASJ95}
\draft
\newcommand{\vol}{}
\newcommand{\no}{}
\newcommand{\lett}{}
\newcommand{\spage}{}
\newcommand{\rdate}{1998 July 10}
\newcommand{\adate}{1999 March 22}

\begin{document}

\title{
Optical and CO Radio Observations of Poor Cluster Zwicky~1615.8+3505
\thanks{
Based on observations made at Okayama Astrophysical Observatory (OAO),
Kiso Observatory (KISO), and Nobeyama Radio Observatory (NRO).
KISO is operated by Institute of Astronomy, School of Science,
The University of Tokyo.
OAO and NRO are branches of the National Astronomical Observatory,
an inter-university research institute operated by the Ministry of
Education, Science, Sports and Culture.}}

\author{ 
Akihiko {\sc Tomita} \\
{\it Department of Earth and Astronomical Sciences,
Faculty of Education,} \\
{\it Wakayama University, Wakayama 640-8510} \\
{\it E-mail (AT): atomita@center.wakayama-u.ac.jp}
\\ [6pt]
Hideo {\sc Maehara} \\
{\it Okayama Astrophysical Observatory,
National Astronomical Observatory,} \\
{\it Kamogata-cho, Okayama 719-0232}
\\ [6pt]
Tsutomu T. {\sc Takeuchi}
\thanks{Research fellows of the Japan Society for the Promotion of
Science.} and Kouichiro {\sc Nakanishi} \\
{\it Department of Astronomy, Faculty of Science,
Kyoto University,} \\
{\it Sakyo-ku, Kyoto 606-8502} \\
and \\
Mareki {\sc Honma}~$^{\dagger}$, Yoshinori {\sc Tutui}~$^{\dagger}$,
and Yoshiaki {\sc Sofue} \\
{\it Institute of Astronomy, School of Science,
The University of Tokyo,} \\
{\it Mitaka, Tokyo 181-8588}}

\abst{

The cluster Zwicky~1615.8+3505 is considered to be a dynamically
young poor cluster.
To investigate the morphology and star-formation activity of galaxies
under the environment of a dynamically young poor cluster, we have
performed $V$, $R$, and $I$ surface photometry, optical low-resolution
spectroscopy, and $^{12}$CO~($J$~=~1~$-$~0) line observations for
member galaxies.
Our data show that more than 90\% of the observed galaxies show
regular morphologies and no star-formation activities, indicating that
the environment does not affect these galaxy properties.
Among sixteen galaxies observed, only NGC~6104 shows a significant
star-formation activity, and shows a distorted morphology, indicating
a tidal interaction.
This galaxy contains double knots, and only one knot possesses Seyfert
activity, though the sizes and luminosities are similar to each other;
we also discuss this feature.}

\kword{
Clusters of galaxies: individual (Zwicky~1615.8+3505) ---
Galaxies: individual (NGC~6104) ---
Poor clusters}

\maketitle
\thispagestyle{headings}

\clearpage

\section{Introduction}

The cluster Zwicky~1615.8+3505 is a poor cluster 
cataloged in CGCG (Catalogue of Galaxies and of Clusters of Galaxies;
Zwicky et al.\ 1961 -- 1968) and not cataloged as a rich cluster in
Abell et al.\ (1989), and is also known as N45-389 in the X-ray field
(e.g., Price et al.\ 1991).
The cluster contains a peculiar galaxy;
NGC~6109 is a well-developed head-tail radio source, B2~1615+35;
Venkatesan et al.\ (1994) have argued that a poor cluster with such
a radio source has a large internal motion of members, which indicates
being dynamically young.
In the bottom-up scenario, a dynamically young poor cluster is an
early stage in the evolution of structures in the Universe.
In the distant Universe, many galaxies with a distorted morphology and
with star-formation activity are observed
(e.g., Williams et al.\ 1996; Smail et al.\ 1997;
Dressler et al.\ 1999).
Although the environmental effects on galaxies have been studied, they
are complicated and not yet clear (e.g., Irwin 1995;
Takeuchi et al.\ 1999).
To investigate the morphology and star-formation activity of galaxies
under the environment of dynamically young poor cluster, we performed
optical and CO radio observations of member galaxies.
In this paper, we also present the basic optical and CO data in order
to discuss such a dynamically young poor cluster.

We describe the observations in section~2.
In section~3 we present the results, and a discussion is given in
section~4.
The characteristics of the cluster are summarized in table~1.

\section{Observations}

We carried out optical surface photometry, long-slit optical
spectroscopy, and $^{12}$CO~($J$~=~1~$-$~0) line observations.
The observed objects are listed in table~2 and the observation log is
given in table~3.

Optical surface photometry in the $V$, $R$, and $I$ bands for thirteen
galaxies was performed in 1995 May at the Kiso Observatory (hereafter
KISO) using the 1.05-m F~3.1 Schmidt telescope equipped with a
single-chip CCD camera at the prime focus.
The CCD chip has 1000 $\times$ 1018 pixels and one pixel size
corresponds to $0.\hspace{-2pt}''752$, giving a field-of-view of
$12.\hspace{-2pt}'5$ $\times$ $12.\hspace{-2pt}'7$.
For all galaxies, we took three frames in each band;
the exposure time of one frame was 20 min.
By combining three frames, we obtained a set of $V$, $R$, and $I$
images with an integration time of 60 min for each galaxy.
In the data reduction, we used IRAF in the usual manner (IRAF is the
software developed in National Optical Astronomy Observatories, USA).
We also used SPIRAL (Hamabe, Ichikawa 1992) in sky-background
subtraction.
To obtain the standard Johnson--Kron--Cousins photometry system, we
converted the observed CCD counts using following equations:

$v = V + c_{V1} + c_{V2} F(z) + c_{V3} (V-R)$,

$r = R + c_{R1} + c_{R2} F(z) + c_{R3} (V-R)$,

$i = I + c_{I1} + c_{I2} F(z) + c_{I3} (V-I)$,

\noindent
where $v$, $r$, and $i$ are the observed magnitudes, defined as
$-2.5$~log~(count)~+26.0 for 300 s-exposure frames;
$V$, $R$, and $I$ are the magnitudes in the standard system, and
$F(z)$ is the air-mass function.
The standard stars were chosen from a list by Landolt (1992).
Nine coefficients $c$ with subscripts were determined every night by
analyzing the standard star frames using the IRAF package APPHOT.
The derived values of the coefficients are given in table~4.
Using the coefficients, we measured the throughput of the telescope,
which is shown in appendix~1.
The seeing size was around $4''$ for all bands and nights.

Long-slit optical spectroscopy for eight galaxies was carried out in
1995 April and May at the Okayama Astrophysical Observatory (hereafter
OAO) using the 1.88-m reflector.
A spectrograph with a grating of 150 grooves mm$^{-1}$ blazed at
5000~\AA\ and a Photometrics CCD of 516 $\times$ 516 pixels was
equipped at the Cassegrain focus.
We binned the spatial direction by two, and resultant one pixel size
corresponds to $1.\hspace{-2pt}''5$.
The dispersion was 4.9~\AA~pixel$^{-1}$, and the wavelength range
covered between 4600 and 7100~\AA.
For each galaxy, we observed one position, putting a slit with a
position angle of 90$^{\circ}$ (east -- west direction), centered on
the galactic center.
The slit length was $4'$, and we obtained a sufficient region to
subtract the sky emissions well.
The exposure time of one frame was 20 min.
Except for galaxy No.~1 (CGCG~1612.2+3500), we observed three times
for each object, and combined these frames;
we then obtained images with an integration time of 60 min.
In the data reduction, we used IRAF in the usual manner.
We also used SNGRED, a semi-automatic reduction package for
spectroscopic data, being operated on IRAF.
The flux was calibrated using the flux-standard star frames (HZ~44 and
Ross~640);
we obtained the relatively flux-calibrated spectral frames.
We also measured the throughput of the telescope (appendix~1).
The seeing size was around $3''$.

$^{12}$CO~($J$~=~1~$-$~0) line observations for six galaxies were
made in 1995 January and March, and 1996 April at the Nobeyama Radio
Observatory (hereafter NRO) using the 45-m telescope.
The FWHM of the beam size was $16''$, which corresponds to 9.5~kpc in
physical size at an assumed angular distance of 122~Mpc (see table~1).
We made on-off switching observations with the on-position centered on
the galaxy centers.
Only for galaxy No.~3 (NGC~6104), we observed another position off the
center.
An acousto-optical spectrometer was used, which had a velocity
coverage of 650~km~s$^{-1}$ with 2048 channels.
The temperature calibration was made with an absorbing chopper at
297~K in front of the receiver.
A reduction system, NEWSTAR at NRO, was used for the data reduction.

\section{Results}

\subsection{Optical Surface Photometry}

\subsubsection{Morphology and size}

Figure~1 shows $R$-band images for thirteen observed galaxies.
All of the frames cover 103 $\times$ 103 pixels or $77'' \times 77''$
field-of-view.
The length of one side of the frame corresponds to 46~kpc in physical
size from the distance of the cluster.

The morphologies of the galaxies were determined using the $R$-band
images (see figure~1).
The second column of table~5 gives the morphology in the de
Vaucouleurs-system and the third column gives the $T$ index, as is in
RC3 (de Vaucouleurs et al.\ 1991) with an error;
inspections of morphology were made independently by two of the
authors (A.T. and T.T.T.), and the error was estimated from the
difference between the two inspections.

Galaxy No.~3 (NGC~6104) is a peculiar galaxy having double knots, and
can not be assigned the Hubble classification.
We named knot~A and knot~B for the nuclear knot and the western knot,
respectively (see figure~1a).
Outer isophotal contours with $\mu_{R}$~=~23.5 to 24~mag~arcsec$^{-2}$
show a water drop-like distorted shape, and a tail feature is seen
from knot~B to the lower right direction in figure~1a.
Except for this one peculiar galaxy, all galaxies have normal
morphologies and are early-type galaxies.

We measured the shapes of the galaxies on the $R$-band images using
the STSDAS package of ISOPHOTE.
In measuring the shapes, we took a level of the $R$-band surface
brightness of $\mu_{R}$~=~23.5~mag~arcsec$^{-2}$ and fitted an
ellipse.
The CCD count for the surface brightness corresponds to about
four-times the rms of the background noise in the $V$- and $R$-band
frames, and about three times in the $I$-band frames.
If taking $B-R = 1.5$, the surface brightness corresponds to
$\mu_{B}$~=~25.0~mag~arcsec$^{-2}$, which has been used in many
previous studies, such as in RC3. 

The fourth, fifth, and sixth columns of table~5 give the major-axis
diameter, ellipticity, and position angle of the major axis for the
observed thirteen galaxies, respectively.
The seeing blurring was not corrected in deriving the values.
The errors were taken from an estimation given in the output through
ISOPHOTE.
Galaxy No.~10 (NGC~6109) was taken twice on different nights and in
different CCD fields-of-view.
The difference in the two independent measurements is well within the
error.
This indicates that our observations and reduction procedures were
successful.

\subsubsection{Magnitudes and colors}

In the following analyses of magnitudes and colors, we used frames
after removing foreground stars using the IRAF task of IMEDIT.
We measured the isophotal magnitudes and colors by integrating the
region within the ellipsoid of $\mu_{R} = 23.5$~mag~arcsec$^{-2}$.
In table~6, the isophotal $V$-band magnitudes ($V_{23.5}$), $V-R$ and
$V-I$ colors [$(V-R)_{23.5}$ and $(V-I)_{23.5}$] are presented in the
second, third, and fourth columns, respectively.
Galaxy No.~10 has two independent frames, and by inspecting the
consistency between two measurements, we estimated the error of the
isophotal magnitudes and colors to be 0.02~mag.
The frame qualities for galaxies Nos.~12 and 14 were relatively more
poor than others, and we estimated the errors for them to be 0.1~mag.

The asymptotic total $V$-band magnitudes ($V_{\rm T}$) were measured
following the descriptions in {\it Photometric Atlas of Northern
Bright Galaxies} (Kodaira et al.\ 1990).
We made growth curves of the $V$-band magnitudes and fitted the
observed growth curves to the templates of the growth curves given in
Kodaira et al.\ (1990).
The templates are given for two categories divided by an axial ratio
at $\mu_{B} = 25.0$~mag~arcsec$^{-2}$ of log~$R_{25} = 0.30$.
We estimated the axial ratio as $R_{25}$~=~$[1/(1-e)]$, where $e$ is
the ellipticity at $\mu_{R} = 23.5$~mag~arcsec$^{-2}$, given in
table~5.
In most cases, we could successfully fit to templates with an
appropriate morphology and axial ratio.
If it was not successful, we used another template which fit the
observed growth curve best.
The fitting was made by two of the authors (A.T. and T.T.T.), and
the results were consistent with each other;
we estimated the error of measurement to be 0.03~mag.
Because the original frames for galaxies Nos.~12 and 14 were poor, the
errors for them were estimated to be 0.1~mag.
The fifth column of table~6 gives the measured $V_{\rm T}$.

The total asymptotic colors in $V-R$ and $V-I$ [$(V-R)_{\rm T}$ and
$(V-I)_{\rm T}$] were measured following the descriptions in Buta,
Crocker (1992).
We used template growth curves in the $B-V$ and $U-B$ colors given in
RC2 (de Vaucouleurs et al.\ 1976) for templates in the $V-R$ and $V-I$
colors, respectively, fixing the $B$-band effective radius
$r_{\rm e}$.
We estimated $r_{\rm e}$ using observed effective radius in the
$V$-band $r_{\rm e}(V)$, and the data of the template growth curves in
the $B-V$ color, depending on the morphology.
We obtained log~$r_{\rm e}/r_{\rm e}(V)$ to be 0.01 for galaxies
earlier than Sa, 0.02 for Sa, 0.03 for Sab and Sb, and 0.04 for Sbc.
The last column of table~5 lists the estimated $r_{\rm e}$;
The Hubble-type morphology for galaxy No.~3 (NGC~6104) is unknown, and
in calculating $r_{\rm e}$ for galaxy No.~3 we assigned $T = 4$, which
is the latest among the other galaxies.
The fitting was made by two of the authors (A.T. and T.T.T.), and the
results were consistent with each other;
we estimated the error of the measurement to be 0.03~mag.
Because the original frames for galaxies Nos.~12 and 14 were poor, the
errors for them were estimated to be 0.1~mag.
The sixth and seventh columns of table~6 give the measured
$(V-R)_{\rm T}$ and $(V-I)_{\rm T}$.

The corrected total magnitudes and colors [$V_{\rm T}^{0}$,
$(V-R)_{\rm T}^{0}$, and $(V-I)_{\rm T}^{0}$] were measured following
the descriptions in Buta and Williams (1995);
the correction considers the extinctions by our Galaxy and the
galaxies themselves and the redshift effect ($k$-correction).
RC3 gives the Galactic extinction in $B$-band $A_{\rm g}$ for seven
galaxies (galaxies Nos.~3, 6, 8, 10, 12, 14, and 15) of 0.00 or
0.01~mag.
We neglected the Galactic extinction in the $V$, $R$, and $I$-bands
toward the region of the cluster.
The color excess in $B-V$ due to internal extinction $E(B-V)$ was
obtained by equation (63) in RC3.
The axial ratio of $R_{25}$ used in the equation was given as
$R_{25} = 1/(1-e)$, where $e$ is the ellipticity given in table~5.
The extinction in the $V$-band $A_{V}$ and the color excesses in
$V-R$ and $V-I$ [$E(V-R)$ and $E(V-I)$] are calculated through three
ratios: $R_{V} = A_{V} / E(B-V)$, $R(V-R) = E(V-R) / E(B-V)$, and
$R(V-I) = E(V-I) / E(B-V)$.
From equation (59) in RC3, we obtained $R_{V} = 3.4$ adopting
$B-V = 1.0$, corresponding to $V-R = 0.6$, using the relation shown in
figure~5 of Buta and Williams (1995).
Using the relation between $R_{V}$ and $R(V-R)$ or $R(V-I)$ given in
Buta and Williams (1995), we obtained $R(V-R) = 0.52$ and
$R(V-I) = 1.20$.
We did not correct the internal extinction for galaxy No.~3
(NGC~6104);
the intrinsic Hubble-type is unknown and the extinction may be small,
because of a small apparent inclination.
Following Buta and Williams (1995), we obtained a $k$-correction by
referring to the data given in Schneider et al.\ (1983);
we took $k(V-R) = 0.05$ and $k(V-I) = 0.05$ for all galaxies.
The last three columns of table~6 give the measured $V_{\rm T}^{0}$,
$(V-R)_{\rm T}^{0}$, and $(V-I)_{\rm T}^{0}$.
The error is estimated to be 0.1~mag, except for galaxies Nos.~12 and
14, 0.2~mag.

Figure~2 shows a $(V-R)_{\rm T}^0$ vs $(V-I)_{\rm T}^0$ diagram for
the thirteen galaxies observed at KISO.
Buta and Williams (1995) gives the standard colors for each galaxy
morphology.
It is found that all of the observed galaxies have intrinsic red
colors, which correspond to those for galaxies earlier than or equal
to Sa.

\subsection{Optical Spectroscopy}

The resultant spectra of the eight observed galaxies are shown in
figure~3.
On the slit, we integrated the spatial direction within the
$\mu_{R} = 20.5$~mag~arcsec$^{-2}$ isophotal position.
The signal count for $\mu_{R} = 20.5$~mag~arcsec$^{-2}$ in our OAO
frames was about twice the background noise.
The sizes of the regions for the integration were 8~pix ($12''$),
10~pix ($15''$), 15~pix ($22.\hspace{-2pt}''5$), 11~pix
($16.\hspace{-2pt}''5$), 10~pix ($15''$), 8~pix ($12''$), 6~pix
($9''$), and 7~pix ($10.\hspace{-2pt}''5$) for galaxies Nos. 1, 2, 3,
7, 10, 12, 14, and 15, respectively ($1''$ corresponds to 0.6~kpc).
Since the quality of spectrum for galaxy No.~1 is relatively poor, we
integrated a region of inner 6~pix for the galaxy.
The sizes of the regions were determined while referring to the KISO
$R$-band images.
If the galaxy did not have the KISO image, we estimated the region for
the integration using equation (1) in appendix~1.

Except for galaxy No.~3 (NGC~6104), the spectra show few signs for
emission lines, and the spectral energy distributions (SED) resemble
those of early-type galaxies (e.g., Kennicutt 1992), which is
consistent with the KISO color analysis mentioned in subsubsection
3.1.2.
The strong sky emission line of the sodium D line caused a spurious
line at about 5900~\AA\ in the reduced spectra, which is serious in
spectra for galaxies Nos.~1, 7, and 14.
The heliocentric radial velocity of galaxy No.~3 was derived to be
$8313 \pm 100$~km~s$^{-1}$ after a correction of the Earth's motion;
this is consistent with value in NED of $8382 \pm 50$~km~s$^{-1}$.

Galaxy No.~3 (NGC~6104) has a complex spatial structure;
the slit position was placed at the bar structure penetrating two
knots.
The region used for figure~3c was divided equally into three (5
pixels in OAO data, $7.\hspace{-2pt}''5$, note that this is larger
than the seeing size of $3''$ at OAO);
from west to east, we call the regions knot~B (western knot),
knot~A (nuclear knot), and Eastern Region, respectively.
Figure~4 shows the spectra of (a) knot~A, (b) knot~B, and (c)
Eastern Region, and eight marks of small vertical lines indicate
expected positions of redshifted emission lines;
from left to right,
H$\beta$, [O\,{\footnotesize III}]$\lambda\lambda4959,5007$,
[N\,{\footnotesize II}]$\lambda6548$, H$\alpha$,
[N\,{\footnotesize II}]$\lambda6583$, and
[S\,{\footnotesize II}]$\lambda\lambda6716,6731$.
The spectrum of knot~A shows a Seyfert feature;
the FWHM of the broad H$\alpha$ line is about 10000~km~s$^{-1}$ and
the equivalent width of the H$\alpha$ line is about 90~\AA.
A broad H$\beta$ line is marginally seen and strong forbidden lines
of [O\,{\footnotesize III}], [N\,{\footnotesize II}], and
[S\,{\footnotesize II}] also indicate the Seyfert characteristics.
At knot~B, the broad H$\alpha$ component and strong forbidden lines
vanish.
Although NGC~6104 has been known as a Seyfert 1 galaxy (e.g., Hewitt,
Burbidge 1991), we found that only knot~A possesses the Seyfert
characteristics.
The Eastern Region has a blue SED;
the continuum is more intense in shorter wavelengths.
Only a sharp H$\alpha$ line is prominent, which resembles the spectra
of the H\,{\footnotesize II} regions, though the equivalent width of
the H$\alpha$ line is only about 10~\AA.
Table~7 summarizes the characteristics of the H$\alpha$ lines in three
regions, and also gives the regional colors derived from the KISO
data;
we took circular regions with a radius of $3.\hspace{-2pt}''75$
(5~pix) in the KISO data.
At knot~A, the H$\alpha$ profile was decomposed into broad and narrow
components by the fitting with two Gaussian curves.
The H$\alpha$ luminosities were calculated from the observed H$\alpha$
surface brightness of the narrow component multiplied by the areas of
the circular regions with a radius of $3.\hspace{-2pt}''75$.

In other galaxies, we estimated the upper limits of the H$\alpha$
fluxes.
Assuming that the observed H$\alpha$ line width of 7~\AA, which is the
case for the H$\alpha$ narrow component in galaxy No.~3, and upper
limit of peak H$\alpha$ emission is three-times the rms of noise, we
calculated the upper limits of the surface brightnesses of the
H$\alpha$ emission.
Multiplying by slit aperture areas for the spectra of figure~3, we
obtained the upper limits of the H$\alpha$ luminosities, though the
slit aperture did not cover the entire galaxy.
Table~8 lists the upper limits;
following Kennicutt (1983), we derived the upper limits of the
star-formation rates from the H$\alpha$ data.

\subsection{CO Observations}

Figure~5 shows the resultant spectra of the $^{12}$CO~($J$~=~1~$-$~0)
line for six observed galaxies.
The observed positions were set to the galactic centers for all
galaxies, except for galaxy No.~3, for which we also observed the
Eastern Region, $7.\hspace{-2pt}''5$ east of the nucleus as well as
knot~A (nuclear knot).
The beam size of $16''$, corresponding to 9.5~kpc, is about half the
sizes of the galaxies listed in the fourth column of table~5;
this means that the observed beams covered most parts of the galaxies.
The center frequency was set to the redshifted
$^{12}$CO~($J$~=~1~$-$~0) frequency, and the redshifts were obtained
from the NED, which are listed in table~2.
The spectra were smoothed every 64 channels, resulting in a velocity
resolution of 21~km~s$^{-1}$.
The rms noise temperatures after smoothing are 3.9, 3.9, 3.3, 3.6,
3.6, 4.7, and 3.5~mK for galaxy No.~3 knot~A, No.~3 Eastern
Region, Nos.~6, 7, 10, 14, and 15, respectively.
The dotted lines in figure~5 indicate twice the rms.
No significant emissions above three-times the rms were detected
for all positions.
A marginal emission feature is seen in the Eastern Region of
galaxy No.~3 (NGC~6104) (see figure~5b) at 17 to 20 channels.
The radial velocity of the H$\alpha$ line at the Eastern Region is
$8189 \pm 100$~km~s$^{-1}$, which is not consistent with the radial
velocity of the CO line, 8428~km~s$^{-1}$, measured from figure~5b;
the feature seems to be noise.

We measured the upper limit of the molecular mass.
The molecular mass $M$(H$_{2}$) and observed CO intensity $I$(CO)
have the relation

$M$(H$_{2}$) = $a$ $\Omega$ $d^{2}$ $I$(CO) / $(1+z)^{3}$,

\noindent
where $a$ is the CO-to-H$_{2}$ conversion factor, and here we take the
Galactic value of 4.5~$\MO$ (K~km~s$^{-1}$~pc$^{2}$)$^{-1}$
(e.g., Sanders et al.\ 1984; Bloemen et al.\ 1986;
note Arimoto et al.\ 1996);
$\Omega$ is a solid angle of the telescope
beam, and we take $16''$ aperture, $4.726 \times 10^{-9}$~str;
$d$ is the luminosity distance, and we take 130~Mpc (see table~1);
and $z$ is the redshift, 0.03148.
Then, we obtain

$M$(H$_{2}$) [$\MO$] = $3.265 \times 10^{8}$ $I$(CO) [K~km~s$^{-1}$].

\noindent
We assumed that a potential emission line has a width of
100~km~s$^{-1}$.
The rms in 100~km~s$^{-1}$-bin, $\sigma_{100}$, was estimated from the
rms in figure~5 (in 21~km~s$^{-1}$-bin), $\sigma$, as
$\sigma_{100} = (21/100)^{1/2} \sigma$.
We took the upper limit of the CO intensity as
$I$(CO)$_{\rm lim} = 3 \times \sigma_{100}$~[K]~
$\times 100$~km~s$^{-1}$.
The upper limits of $M$(H$_{2}$) within a central 9.5~kpc-diameter
aperture for observed six galaxies are summarized in table~8.

Though galaxy No.~3 (NGC~6104) is not cataloged in {\it Atlas of
Peculiar Galaxies} by Arp (1966), it is an Arp-like galaxy.
Sofue et al.\ (1993) studied the CO content of Arp's galaxies.
Galaxy No.~3 is an $IRAS$ source of F16146+3549 and its
far-infrared luminosity is log~$L$(FIR)~[$\LO$] =~10.5.
From Sofue et al.\ (1993), for an Arp galaxy with
log~$L$(FIR)~[$\LO$] =~10.5, log~$M$(H$_{2}$)~[$\MO$] is 8.5 to 9.5.
This is larger than the upper limit obtained by us for galaxy No.~3.
Maiolino et al.\ (1997) reported a molecular mass of about
$10^{10}$~$\MO$ in NGC~6104 (galaxy No.~3) in a beam with a
diameter of $1'$;
their beam area was about ten-times ours.
The molecular clouds may exist in an outer region of the galaxy.
We should note that the CO spectrum of NGC~6104 by
Maiolino et al.\ (1997) has a poor signal-to-noise ratio, and they
wrote that the baseline fitting was relatively poor.

\section{Discussion}

\subsection{Effect of Cluster Environment}

Galaxy No.~10 (NGC~6109) is known as a well-developed head-tail radio
source, B2~1615+35, and has been studied in the radio continuum and
X-ray regions (Ekers et al.\ 1978; Burns, Gregory 1982;
Burns et al.\ 1987; Feretti et al.\ 1995).
Head-tail radio sources are usually found in rich clusters with a high
density of intracluster medium (ICM) and a high speed of member
galaxies relative to ICM, which make a high ram pressure.
Venkatesan et al.\ (1994) have argued that a poor cluster with
developed head-tail radio sources must have a high relative speed,
which means that the cluster is dynamically young.
Ulrich (1978) also discussed that this cluster consists of two
subclusters.

Out of thirteen galaxies observed at KISO, all galaxies except for
galaxy No.~3 have regular morphologies.
Out of the sixteen galaxies observed either at KISO or OAO or NRO, all
galaxies except for galaxy No.~3 do not show any star-formation
activities.
From the KISO data, the colors of the fifteen galaxies are what can be
expected from their morphologies (see figure~2), and there is no
peculiar surface-color distribution in each galaxy.
These facts indicate that in this cluster the environment of a
dynamically young poor cluster does not affect the morphologies and
star-formation activities of the member galaxies.

Vigroux et al.\ (1989) discussed that star-formation activities of
member galaxies are activated at early stages of cluster dynamical
evolution, by showing galaxies with recent star-formation activities
in a poor cluster, Pegasus~I.
Though the X-ray luminosity of Pegasus~I is similar to that of cluster
Zwicky~1615.8+3505 (Giovanelli, Haynes 1985), five out of ten
early-type galaxies show recent star formation in Pegasus~I, and this
is in contrast to our result in Zwicky~1615.8+3505.
The star-formation activity may be independent of the cluster
dynamical stage (see also Tomita et al.\ 1996).
The compilation of more data of star formation properties in various
cluster galaxies is necessary to confirm the above-mentioned result.

\subsection{A Peculiar Galaxy, NGC~6104 (galaxy No.~3)}

Only galaxy No.~3 (NGC~6104) shows significant star-formation activity
among the observed galaxies, and has a star-formation rate of about
0.1~$\MO$~yr$^{-1}$ (see table~8).
The galaxy has the bluest total color among the thirteen galaxies
observed at KISO, indicating present star-formation activity as well.
Only this galaxy has a peculiar morphology among the observed
galaxies, as shown in figure~1a.
The tail feature and distorted outer isophotal contours indicate that
this galaxy is under a tidal interaction, or possibly a merger.

Both knot~A and knot~B have similar radial velocities to each other,
which means that the galaxy surely has double knots, not by a chance
alignment;
the velocity difference between the two knots is 170~km~s$^{-1}$.
Only one (knot~A) of the double knots shows Seyfert activity, though
the sizes and luminosities of the two knots are similar to each other;
the size is $7.\hspace{-2pt}''5$ or about 4~kpc in diameter (note that
this is larger than the seeing size of $4''$ at KISO), and in an
aperture with a diameter of $7.\hspace{-2pt}''5$, $I$-band magnitudes
are 14.8 and 15.1 for knot~A and knot~B, respectively, and thus,
$I$-band luminosities differ by a factor of 1.3.
Knot~B may be a giant H\,{\footnotesize II} region or hot spot at
the end of the bar structure induced by the tidal interaction.
It is also possible that knot~B is a remnant of a possible merger.
According to Sanders et al.\ (1988), merger events would generate AGN
activity accompanying the starburst.
If the two knots are remnants of pre-existing galaxies, a question
arise as to why only one knot possesses Seyfert activity.
The size and luminosity are similar to each other, and considering
that knot~A contains the Seyfert, knot~A may not be more massive than
knot~B.
In order to study the above issue further, we need higher spatial
resolution imaging observations so as to investigate whether the
double knots are merger remnants.

\vspace{1pc}


We would like to thank Mamoru Sait\={o} and Nagako Miyauchi-Isobe
for valuable discussion.
We are grateful to thank the OAO, KISO, and NRO staff members for
their hospitality during our stay.
Franz Sch\"{o}niger kindly helped us in the observations at NRO.
T.T.T., M.H., and Y.T.\ acknowledge the Research Fellowship of the
Japan Society for the Promotion of Science for Young Scientists.
This research also made use of the NASA/IPAC Extragalactic Database
(NED), which is operated by the Jet Propulsion Laboratory, Caltech,
under contact with the National Aeronautics and Space Administration.

\section*{Appendix 1.\ Throughputs of Telescopes at KISO and OAO}

The throughput of the KISO 1.05-m Schmidt telescope, without
considering air absorption, is derived as follows.
The output electron number on the CCD for a 0.0-mag object with 300
s-exposure $N_{\rm e}$ is given by log~$N_{\rm e}'$~=~10.4~$-$~0.4
$c_{(V,R,I)1}$ (see table~4 for coefficients),
$N_{\rm e}$~=~0.88~$N_{\rm e}'$ (AD conversion of the CCD camera was
0.88 electrons~ADU$^{-1}$).
In calculating the expected photon number from objects, we referred to
the data given by Fukugita et al.\ (1995);
the band flux for a 0.0-mag object was calculated to be
$f_{\lambda, {\rm eff}}^{\rm Vega}$ (converted for 0.0 mag)
$\times$ FWHM, and the energy of a photon for a band was calculated as
$hc/\lambda_{\rm eff}$.
We took the aperture of a complete circle with a diameter of 1.05 m,
and then obtained throughputs of 29, 18, and 14\% in the $V$, $R$, and
$I$ bands, respectively. 

The throughput of the OAO 1.88-m telescope, including air absorption,
was derived as follows.
We obtained a relation between the count rate of the spectrograph CCD
at 6000~\AA\ on one pixel $C$ [count~s$^{-1}$], and the $R$-band
surface brightness measured from KISO images for a spectrograph
condition of (1) a dispersion of 4.9~\AA~pix$^{-1}$, (2) a slit width
of 0.30~mm, corresponding to $1.\hspace{-2pt}''857$, (3) a spatial
resolution of $1.\hspace{-2pt}''5$~pix$^{-1}$, and (4) a CCD gain of
2.35~electrons~ADU$^{-1}$ (the gain value on the SNG PC is 90),

\begin{equation}
\mu_{R} = 15.5 - 2.5~{\rm log}~C.
\end{equation}

\noindent
We took the aperture of a complete circle with a diameter of 1.88~m;
then, the throughput of the telescope-spectrograph system at 6000~\AA,
including air absorption, turned out to be about 1.5\%.

\section*{Appendix 2.\ KUG Catalog in This Field}

The cluster was also noticed as the KUG-rich cluster.
The KUGs, Kiso Ultraviolet-excess Galaxies, are a UV-excess galaxy
sample surveyed by means of plates taken by the Kiso Observatory
1.05-m Schmidt telescope, the compiled catalog of which was given by
Takase, Miyauchi-Isobe (1993).
Tomita et al.\ (1997) made a statistical analysis of the general
properties of the KUGs.
Takeuchi et al.\ (1999) analyzed the KUG fraction, the fraction of
being KUGs among all galaxies, and found regions where the KUG
fractions are exceptionally high; the cluster region was included.
However, the survey in the cluster region was found to be erroneous,
and Miyauchi-Isobe et al.\ (1997) revised the catalog.

\clearpage

\section*{References}

\re
Abell G.O., Corwin H.G.\ Jr, Olowin R.P.\ 1989,
ApJS 70, 1

\re
Arimoto N., Sofue Y., Tsujimoto Y.\ 1996,
PASJ 48, 275

\re
Arp H.\ 1966,
Atlas of Peculiar Galaxies
(California Institute of Technology, California)

\re
Bloemen J.B.G.M., Strong A.W., Blitz L., Cohen R.S., Dame T.M.,
Grabelsky D.A., Hermsen W., Lebrun F. et al.\ 1986,
A\&A 154, 25

\re
Burns J.O., Gregory S.A.\ 1982,
AJ 87, 1245

\re
Burns J.O., Hanisch R.J., White R.A., Nelson E.R., Morrisette K.A.,
Moody J.W.\ 1987,
AJ 94, 587

\re
Buta R., Crocker D.A.\ 1992,
AJ 103, 1804

\re
Buta R., Williams K.L.\ 1995,
AJ 109, 543

\re
de Vaucouleurs G., de Vaucouleurs A., Corwin H.G.\ Jr 1976,
Second Reference Catalogue of Bright Galaxies
(University of Texas Press, Austin) (RC2)

\re
de Vaucouleurs G., de Vaucouleurs A., Corwin H.G.\ Jr, 
Buta R.J., Paturel G., Fouqu\'{e} P.\ 1991,
Third Reference Catalogue of Bright Galaxies
(Springer-Verlag, New York) (RC3)

\re
Dressler A., Smail I., Poggianti B., Butcher H., Couch W., Ellis R.,
Oemler A.\ Jr 1999,
ApJS in press

\re
Ekers R.D., Fanti R., Lari C., Ulrich M.-H.\ 1978,
A\&A 69, 253

\re
Feretti L., Fanti R., Parma P., Massaglia S., Trussoni E.,
Brinkmann W.\ 1995,
A\&A 298, 699

\re
Fukugita M., Shimasaku K., Ichikawa T.\ 1995,
PASP 107, 945

\re
Giovanelli R., Haynes M.P.\ 1985,
ApJ 292, 404

\re
Hamabe M., Ichikawa S.\ 1992,
in Proc.\ of Astronomical Data Analysis Software and System~I,
ed D.M.\ Worrall, C.\ Biemesderfer, J.\ Barnes,
ASP Conf.\ Ser.\ 25, p325

\re
Hewitt A., Burbidge G.\ 1991,
ApJS 75, 297

\re
Horaguchi T., Ichikawa S., Yoshida M., Yoshida S., Hamabe M.\ 1994,
Publ.\ Natl.\ Astron.\ Obs.\ Jpn 4, 1

\re
Irwin J.A.\ 1995,
PASP 107, 715

\re
Kennicutt R.C.\ Jr 1983,
ApJ 272, 54

\re
Kennicutt R.C.\ Jr 1992,
ApJS 79, 255

\re
Kodaira K., Okamura S., Ichikawa S.\ 1990,
Photometric Atlas of Northern Bright Galaxies
(University of Tokyo Press, Tokyo)

\re
Landolt A.U.\ 1992,
AJ 104, 340

\re
Maiolino R., Ruiz M., Rieke G.H., Papadopoulos P.\ 1997,
ApJ 485, 552

\re
Miyauchi-Isobe N., Takase B., Maehara H.\ 1997,
Publ.\ Natl.\ Astron.\ Obs.\ Japan 4, 153

\re
Price R., Burns J.O., Duric N., Newberry M.V.\ 1991,
AJ 102, 14

\re
Sanders D.B., Solomon P.M., Scoville N.Z.\ 1984,
ApJ 276, 182

\re
Sanders D.B., Soifer B.T., Elias J.H., Madore B.F., Matthews K.,
Neugebauer G., Scoville N.Z.\ 1988,
ApJ 325, 74

\re
Schneider D.P., Gunn J.E., Hoessel J.G.\ 1983,
ApJ 264, 337

\re
Smail I., Dressler A., Couch W.J., Ellis R.S., Oemler A.\ Jr,
Butcher H., Sharples R.M.\ 1997,
ApJS 110, 213

\re
Sofue Y., Wakamatsu K., Taniguchi Y., Nakai N.\ 1993,
PASJ 45, 43

\re
Takase B., Miyauchi-Isobe N.\ 1993,
Publ.\ Natl.\ Astron.\ Obs.\ Japan 3, 169

\re
Takata T., Ichikawa S., Horaguchi T., Yoshida S., Yoshida M.,
Ito T., Nishihara E., Hamabe M.\ 1995,
Publ.\ Natl.\ Astron.\ Obs.\ Jpn 4, 9

\re
Takeuchi T.T., Nakanishi K., Ishii T.T., Sait\={o} M., Tomita A.,
Iwata I.\ 1999,
ApJS 121, in press

\re
Tomita A., Nakamura F.E., Takata T., Nakanishi K., Takeuchi T.,
Ohta K., Yamada T.\ 1996,
AJ 111, 42

\re
Tomita A., Takeuchi T.T., Usui T., Sait\={o} M.\ 1997,
AJ 114, 1758

\re
Ulrich M.-H.\ 1978,
ApJ 221, 422

\re
Vigroux L., Boulade O., Rose J.A.\ 1989,
AJ 98, 2044

\re
Venkatesan T.C.A., Batuski D.J., Hanisch R.J., Burns J.O.\ 1994,
ApJ 436, 67

\re
Williams R.E., Blacker B., Dickinson M., Van Dyke Dixson W.,
Ferguson H.C., Fruchter A.S., Giavalisco M., Gilliland R.L.
et al.\ 1996,
AJ 112, 1335

\re
Zwicky F., Herzog E., Kowal C.T., Wild P., Karpowicz M.\ 1961 -- 1968,
Catalogue of Galaxies and of Clusters of Galaxies
(California Institute of Technology, Pasadena) (CGCG)

\clearpage

\noindent
Figure Captions

\re
Fig. 1.~
$R$-band images for thirteen observed galaxies at KISO.
North is top and east is left.
The length of one side of the frame is 103 pixel or 77$''$, which
corresponds to 46~kpc in physical size.
(a)~Galaxy No.~3, NGC~6104.
The two knots, knots A and B, are indicated.
(b)~Galaxy No.~4.
(c)~Galaxy No.~5, CGCG~1615.0+3550.
(d)~Galaxy No.~6, NGC~6105.
(e)~Galaxy No.~8, NGC~6108.
(f)~Galaxy No.~9.
(g)~Galaxy No.~10, NGC~6109.
(h)~Galaxy No.~11, NGC~6110.
(i)~Galaxy No.~12, NGC~6112.
(j)~Galaxy No.~13, CGCG~1616.1+3523.
(k)~Galaxy No.~14, NGC~6114.
(l)~Galaxy No.~15, NGC~6116.
(m)~Galaxy No.~16, CGCG~1617.2+3510.

\re
Fig. 2.~
The $(V-R)_{\rm T}^{0}$ vs $(V-I)_{\rm T}^{0}$ diagram for thirteen
galaxies observed at KISO.
The mean color in each morphological type by Buta, Williams (1995) is
signed.

\re
Fig. 3.~
Spectra of eight galaxies observed at OAO.
(a)~Galaxy No.~1, CGCG~1612.2+3500.
(b)~Galaxy No.~2, NGC~6097.
(c)~Galaxy No.~3, NGC~6104.
(d)~Galaxy No.~7, NGC~6107.
(e)~Galaxy No.~10, NGC~6109.
(f)~Galaxy No.~12, NGC~6112.
(g)~Galaxy No.~14, NGC~6114.
(h)~Galaxy No.~15, NGC~6116.

\re
Fig. 4.~
Spectra for three sub-regions of galaxy No.~3 (NGC~6104).
(a) Knot~A (nuclear knot), (b) knot~B (western knot), and (c) Eastern
Region.
The eight marks of small vertical lines indicate the expected
positions of redshifted emission lines;
from left to right, H$\beta$, [O\,{\footnotesize III}] doublet,
H$\alpha$ and [N\,{\footnotesize II}] doublet, and
[S\,{\footnotesize II}] doublet.

\re
Fig. 5.~
CO spectrum of six galaxies observed at NRO.
Smoothing of original 64 channels was done, and one channel in the
figure corresponds to 21~km~s$^{-1}$.
The galaxy No.~3 (NGC~6104) was observed at two positions, knot~A
(nuclear knot) and Eastern Region, and other galaxies were observed at
galaxy centers.
(a)~Galaxy No.~3, NGC~6104 knot~A.
(b)~Galaxy No.~3, NGC~6104 Eastern Region.
(c)~Galaxy No.~6, NGC~6105.
(d)~Galaxy No.~7, NGC~6107.
(e)~Galaxy No.~10, NGC~6109.
(f)~Galaxy No.~14, NGC~6114.
(g)~Galaxy No.~15, NGC~6116.

\clearpage


\begin{table*}[t]
\small
\begin{center}
Table~1.\hspace{4pt}Characteristics of the cluster.\\
\end{center}
\vspace{6pt}
 \begin{tabular*}{\textwidth}
  {@{\hspace{\tabcolsep}\extracolsep{\fill}}lrcc}
 \hline\hline\\
 & & Reference & Remark \\
 $\alpha$ (1950)                  &
  $16^{\rm h}15^{\rm m}\hspace{-5pt}.\hspace{2pt}8$ & 1 &   \\
 $\delta$ (1950)                  &
  $35^{\circ}05'$                                   & 1 &   \\
 Mean radial velocity             &
  9438~km~s$^{-1}$                                  & 2 & A \\
 Radial velocity dispersion       &
  584~km~s$^{-1}$                                   & 2 & A \\
 Diameter                         &
  4~Mpc                                             & 1 & B \\
 Rood--Sastry type                 &
  F                                                 & 3 &   \\
 Temperature of the hot gas       &
  2.1~keV                                           & 4 & C \\
 Metal abundance of the hot gas   &
  0.1~$Z_\solar $                                   & 4 & C \\
 X-ray luminosity                 &
                                                    &   &   \\
 \, \, $ROSAT$ (0.1 -- 2.4~keV)   &
  $0.9 \times 10^{43}$ erg~s$^{-1}$                 & 4 & D \\
 \, \, $Einstein~Observatory$ (0.5 -- 4.5~keV)&
  $1.1 \times 10^{43}$ erg~s$^{-1}$                 & 3 & D \\
 \hline\\
 \end{tabular*}

References

1.~Zwicky et al.\ (1966) (Volume~III, p84),
2.~Ulrich (1978),
3.~Price et al.\ (1991),
4.~Feretti et al.\ (1995)

Notes

A:~
Ulrich (1978) reported that the cluster possibly consists of two
different subclusters in the same line of sight, subcluster A and B.
The subcluster A has mean radial velocity of 9124~km~s$^{-1}$ and
velocity dispersion of 326~km~s$^{-1}$.
The subcluster B has 10266~km~s$^{-1}$ and 237~km~s$^{-1}$,
respectively.
In the table, we tabulate the values for the case treated as one
cluster, A+B.

B:~
Distance is derived from the mean radial velocity ($v$) and the Hubble
constant of $H_{0}$ = 75~km~s$^{-1}$Mpc$^{-1}$.
We take the luminosity distance, $(v/H_{0})(1+z)$, and the angular
distance, $(v/H_{0})(1+z)^{-1}$, to be 130 and 122~Mpc, respectively,
where $z$ is a redshift of 0.03148.
The size is derived from the apparent size of 2$^{\circ}$ described in
CGCG.

C:~
Note that these data are based on $ROSAT$ observations.

D:~
Luminosity values are proportional to $H_{0}^{-2}$.
Here, we put $H_{0}$ = 75~km~s$^{-1}$Mpc$^{-1}$.

\end{table*}

\clearpage


\begin{table*}[t]
\small
\begin{center}
Table~2.\hspace{4pt}List of objects observed.\\
\end{center}
\vspace{6pt}
 \begin{tabular*}{\textwidth}
  {@{\hspace{\tabcolsep}\extracolsep{\fill}}rccccc}
 \hline\hline\\
 No. & CGCG & NGC  &
 $\alpha(1950)$ & $\delta(1950)$ & $v$ $^{\ast}$ \\
     & name & name &
 [~~$^{\rm h}$~~~$^{\rm m}$~~~$^{\rm s}$] &
 [~~$^{\circ}$~~~$'$~~~$''$] & [km~s$^{-1}$] \\
 \hline\\
 1   & 1612.2+3500 &      & 16 12 14.3 & 34 59 44 & 9807 \\
 2   & 1612.5+3514 & 6097 & 16 12 34.6 & 35 14 01 & 9963 \\
 3   & 1614.7+3550 & 6104 & 16 14 40.1 & 35 49 50 & 8382 \\
 4   &             &      & 16 14 54.7 & 35 24 25 &      \\
 5   & 1615.0+3550 &      & 16 14 59.4 & 35 49 27 & 8177 \\
 6   & 1615.2+3500 & 6105 & 16 15 17.5 & 35 00 02 & 8654 \\
 7   & 1615.4+3501 & 6107 & 16 15 28.4 & 35 01 23 & 9191 \\
 8   & 1615.5+3515 & 6108 & 16 15 34.2 & 35 15 25 & 9116 \\
 9   &             &      & 16 15 44.1 & 35 15 19 &      \\
 10  & 1615.7+3507 & 6109 & 16 15 49.0 & 35 07 30 & 8857 \\
 11  & 1615.8+3513 & 6110 & 16 15 52.5 & 35 12 28 & 9112 \\
 12  & 1616.1+3514 & 6112 & 16 16 09.2 & 35 13 52 & 9319 \\
 13  & 1616.1+3523 &      & 16 16 10.0 & 35 23 12 & 8928 \\
 14  & 1616.5+3518 & 6114 & 16 16 32.3 & 35 17 40 & 8691 \\
 15  & 1617.0+3517 & 6116 & 16 17 03.4 & 35 16 25 & 8800 \\
 16  & 1617.2+3510 &      & 16 17 13.9 & 35 09 06 & 9454 \\    
 \hline\\
 \end{tabular*}

$^{\ast}$~Heliocentric radial velocity taken from NED.

\end{table*}

\clearpage


\begin{table*}[t]
\small
\begin{center}
Table~3.\hspace{4pt}Log of observations.\\
\end{center}
\vspace{6pt}
 \begin{tabular*}{\textwidth}
  {@{\hspace{\tabcolsep}\extracolsep{\fill}}rcccc}
 \hline\hline\\
 No. $^{\ast}$ & KISO $^{\ddagger}$ & OAO $^{\ddagger}$ &
 \multicolumn{2}{c}{NRO} \\
       & Date        & Date   & Date $^{\S}$    & Exp.Time\\
       & (1995)      & (1995) &                    &  [s] \\
 \hline\\
 1~~~  &             & May 7  &                    &      \\
 2~~~  &             & May 5  &                    &      \\
 3~~~  & May 31      & Apr 26 & Jan 15, 23, 25     & 10120\\
~~~$^{\dagger}$&     &        & Apr 11, 12, 20, 22 & 12600\\
 4~~~  & Jun 1       &        &                    &      \\
 5~~~  & May 31      &        &                    &      \\
 6~~~  & May 26      &        & Apr 13, 14         & 15220\\
 7~~~  &             & Apr 26 & Jan 16, Mar 8      & 16880\\
 8~~~  & May 27      &        &                    &      \\
 9~~~  & May 27      &        &                    &      \\
10~~~  & May 26      & Apr 26 & Mar 9              & ~9660\\
11~~~  & May 27      &        &                    &      \\
12~~~  & May 27      & Apr 26 &                    &      \\
13~~~  &May 31, Jun 1&        &                    &      \\
14~~~  & May 27      & May 6  & Apr 15, 17         & 10520\\
15~~~  & May 31      & May 5  & Mar 10             & ~9080\\
16~~~  & Mar 31      &        &                    &      \\
 \hline\\
 \end{tabular*}

\end{table*}

\clearpage


\begin{table*}

$^{\ast}$~The number is in the same order as in table~2.

$^{\dagger}$~The observed position is the Eastern Region,
off the center.

$^{\ddagger}$~You can see the quick look of the raw data of our OAO and KISO
observations through Mitaka-Okayama-Kiso data Archival system
(MOKA);
see http://www.moka.nao.ac.jp/.
The MOKA is operated by Astronomical Data Analysis Center,
Okayama Astrophysical Observatory (National Astronomical Observatory
of Japan), and Kiso Observatory (University of Tokyo) in cooperation
with the Japanese Association Information Processing in Astronomy
(Horaguchi et al.\ 1994; Takata et al.\ 1995).

$^{\S}$~In 1995 for January and March runs,
in 1996 for April run.

\end{table*}

\clearpage


\begin{table*}[t]
\small
\begin{center}
Table~4.\hspace{4pt}Coefficients of the photometry.\\
\end{center}
\vspace{6pt}
 \begin{tabular*}{\textwidth}
  {@{\hspace{\tabcolsep}\extracolsep{\fill}}crrrr}
 \hline\hline\\
 Date (1995) & May 26 & May 27    & May 31    & June 1    \\
 \hline\\
 $c_{V1}$ & $-$3.6465 & $-$3.6261 & $-$3.6626 & $-$3.6565 \\
 $c_{V2}$ &    0.1993 &    0.3156 &    0.3282 &    0.2550 \\
 $c_{V3}$ & $-$0.1160 & $-$0.1280 & $-$0.1145 & $-$0.0833 \\
 \hline
 $c_{R1}$ & $-$3.4458 & $-$3.3795 & $-$3.4591 & $-$3.4583 \\
 $c_{R2}$ &    0.1280 &    0.2098 &    0.2601 &    0.1815 \\
 $c_{R3}$ & $-$0.0760 & $-$0.1192 & $-$0.0954 & $-$0.0535 \\
 \hline
 $c_{I1}$ & $-$2.6397 & $-$2.6480 & $-$2.6926 & $-$2.6429 \\
 $c_{I2}$ &    0.0962 &    0.2103 &    0.2145 &    0.1271 \\
 $c_{I3}$ &    0.0425 &    0.0418 &    0.0716 &    0.0614 \\
 \hline
 \end{tabular*}
\end{table*}

\clearpage


\begin{table*}[t]
\small
\begin{center}
Table~5.\hspace{4pt}Shapes of galaxies observed at KISO.\\
\end{center}
\vspace{6pt}
 \begin{tabular*}{\textwidth}
  {@{\hspace{\tabcolsep}\extracolsep{\fill}}cccccccc}
 \hline\hline\\
No.&Morphology&$T$ index&Diameter&Ellipticity&
 P.A.&$r_{\rm e}$$^{\ast}$\\
    &      &          & [arcsec] &           & [degree] & [arcsec] \\
 \hline\\
 ~3 &S(r)pec/Pec& 99? & 29$\pm$1 & 0.18$\pm$0.07 & ~65$\pm$15 & 13 \\
 ~4 & Sbc  & ~$4\pm$1 & 24$\pm$1 & 0.68$\pm$0.03 & 135$\pm$~1 & 11 \\
 ~5 & E2   & $-5\pm$0 & 18$\pm$1 & 0.15$\pm$0.06 & ~90$\pm$13 & ~6 \\
 ~6 & SX0  & $-2\pm$0 & 24$\pm$1 & 0.15$\pm$0.1~ & 110$\pm$20 & 15 \\
 ~8 & SBab & ~2$\pm$1 & 30$\pm$1 & 0.22$\pm$0.07 & ~80$\pm$10 & 15 \\
 ~9 & S0   & $-2\pm$0 & 17$\pm$1 & 0.47$\pm$0.03 & ~75$\pm$~3 & ~5 \\
 10 & E1   & $-5\pm$0 & 35$\pm$1 & 0.1~$\pm$0.05 & ~60$\pm$20 & 33 \\
 11 & Sab  & ~2$\pm$1 & 20$\pm$1 & 0.46$\pm$0.03 & 110$\pm$~3 & ~6 \\
 12 & E1   & $-5\pm$0 & 28$\pm$1 & 0.05$\pm$0.05 & ~90$\pm$40 & 15 \\
 13 & Sb   & ~3$\pm$1 & 28$\pm$1 & 0.55$\pm$0.03 & ~~0$\pm$~2 & 10 \\
 14 & S0a  & ~0$\pm$1 & 30$\pm$1 & 0.4~$\pm$0.05 & 110$\pm$10 & 14 \\
 15 & SAab & ~4$\pm$1 & 32$\pm$1 & 0.42$\pm$0.04 & ~16$\pm$~3 & 14 \\
 16 & Sa   & ~1$\pm$1 & 20$\pm$1 & 0.38$\pm$0.03 & ~77$\pm$~3 & ~8 \\
 \hline\\
 \end{tabular*}

$^{\ast}$~
Estimated effective radius in $B$-band.
See text in subsubsection~3.1.2.

\end{table*}

\clearpage


\begin{table*}[t]
\small
\begin{center}
Table~6.\hspace{4pt}Colors and magnitudes of galaxies observed at
 KISO.\\
\end{center}
\vspace{6pt}
 \begin{tabular*}{175mm}
  {@{\hspace{\tabcolsep}\extracolsep{\fill}}cccccccccc}
 \hline\hline\\
 No. & \multicolumn{3}{c}{Isophotal $^{\ast}$}           &
       \multicolumn{3}{c}{Total $^{\dagger}$}            &
       \multicolumn{3}{c}{Corrected total $^{\ddagger}$} \\
 \hline
     & $V_{23.5}$     & $(V-R)_{23.5}$      & $(V-I)_{23.5}$      &
       $V_{\rm T}$    & $(V-R)_{\rm T}$     & $(V-I)_{\rm T}$     &
       $V_{\rm T}^0$  & $(V-R)_{\rm T}^{0}$ & $(V-I)_{\rm T}^{0}$ \\
 \hline
~3&13.642& 0.530 & 1.053 & 13.50 & 0.53 & 1.05 & 13.42 & 0.48 & 1.00\\
~4&15.456& 0.667 & 1.341 & 15.10 & 0.62 & 1.20 & 14.47 & 0.49 & 0.96\\
~5&14.567& 0.595 & 1.209 & 14.39 & 0.59 & 1.21 & 14.31 & 0.54 & 1.16\\
~6&14.355& 0.619 & 1.264 & 13.86 & 0.61 & 1.26 & 13.78 & 0.56 & 1.21\\
~8&14.155& 0.603 & 1.217 & 13.87 & 0.60 & 1.22 & 13.67 & 0.53 & 1.13\\
~9&14.995& 0.629 & 1.274 & 14.93 & 0.64 & 1.29 & 14.85 & 0.59 & 1.24\\
10&13.642& 0.612 & 1.233 & 12.91 & 0.60 & 1.20 & 12.83 & 0.55 & 1.15\\
11&14.739& 0.600 & 1.181 & 14.67 & 0.60 & 1.18 & 14.29 & 0.50 & 1.02\\
12&14.01~& 0.61~ & 1.22~ & 13.61 & 0.61 & 1.22 & 13.53 & 0.56 & 1.17\\
13&14.717& 0.623 & 1.275 & 14.52 & 0.60 & 1.24 & 14.03 & 0.49 & 1.04\\
14&14.31~& 0.60~ & 1.18~ & 13.93 & 0.61 & 1.19 & 13.73 & 0.54 & 1.10\\
15&14.095& 0.602 & 1.199 & 13.86 & 0.58 & 1.15 & 13.51 & 0.49 & 1.01\\
16&15.000& 0.604 & 1.209 & 14.88 & 0.59 & 1.20 & 14.62 & 0.51 & 1.09\\
 \hline
 \end{tabular*}

\vspace{2pc}

$^{\ast}$~
The $V$-band magnitude and $V-R$ and $V-I$ colors within an isophotal
ellipsoid of $\mu_{R} = 23.5$~mag~arcsec$^{-2}$.
The errors of measurements were estimated to be 0.02~mag for the
magnitudes and colors, except for galaxies Nos.~12 and 14, the errors
of which were estimated to be 0.1~mag.

$^{\dagger}$~
The asymptotic total $V$-band magnitude and $V-R$ and $V-I$ colors.
The errors of measurements were estimated to be 0.03~mag for the
magnitudes and colors, except for galaxies Nos.~12 and 14, the errors
of which were estimated to be 0.1~mag.

$^{\ddagger}$~
The asymptotic total $V$-band magnitude and $V-R$ and $V-I$ colors
correcting extinction and redshift effects.
The errors of measurements were estimated to be 0.1~mag for the
magnitudes and colors, except for galaxies Nos.~12 and 14, the errors
of which were estimated to be 0.2~mag.

\end{table*}

\clearpage


\begin{table*}[t]
\small
\begin{center}
Table~7.\hspace{4pt}
Characteristics of the H$\alpha$ emission in galaxy No.~3.\\
\end{center}
\vspace{6pt}
 \begin{tabular*}{\textwidth}
  {@{\hspace{\tabcolsep}\extracolsep{\fill}}cccccccc}
 \hline\hline\\
 Region   & \multicolumn{2}{c}{Broad H$\alpha$} &
            \multicolumn{3}{c}{Narrow H$\alpha$} &
          $V-R$ $^{\S}$ & $V-I$ $^{\S}$ \\
          & EW & FWHM $^{\ast}$ &
            EW & $cz$ $^{\dagger}$ & $L$(H$\alpha$) $^{\ddagger}$ &
          & \\
          & [\AA] & [km~s$^{-1}$] &
            [\AA] & [km~s$^{-1}$] & [$10^{40}$~erg~s$^{-1}$] &
          & \\
 \hline\\
 Knot~A    & 92.5 & 10370 &  6.7 & 8306 & 2.3 & 0.65 & 1.27 \\
 Knot~B    &  $-$ &  $-$  &  5.7 & 8476 & 1.7 & 0.57 & 1.10 \\
 E. Region &  $-$ &  $-$  & 10.3 & 8189 & 1.8 & 0.55 & 1.09 \\
 \hline\\
 \end{tabular*}

$^{\ast}$~
The instrumental broadening is corrected for the FWHM of the broad
component shown here, though the correction is very small.
The apparent line width of the narrow component (about 7~\AA) is
almost due to the instrumental broadening.

$^{\dagger}$~
Heliocentric radial velocity ($cz$).
The error of measurement is $\pm$100~km~s$^{-1}$.

$^{\ddagger}$~
These values were derived as follows:
observed surface brightness of the H$\alpha$ emission multiplied by
areas of circular regions with radius of $3.\hspace{-2pt}''75$
(2.5~pix).

$^{\S}$~
These are apparent colors without correction of extinctions.
The color in each region was measured using the data from the KISO
image.

\end{table*}

\clearpage


\begin{table*}[t]
\small
\begin{center}
Table~8.\hspace{4pt}Derived parameters for star-formation activity.\\
\end{center}
\vspace{6pt}
 \begin{tabular*}{\textwidth}
  {@{\hspace{\tabcolsep}\extracolsep{\fill}}cccc}
 \hline\hline\\
 No.& $L$(H$\alpha$) & SFR & $M$(H$_{2}$) \\
    & [$10^{39}$~erg~s$^{-1}$]
    & [$10^{-3}$~$\MO$~yr$^{-1}$]
    & [$10^{8}$~$\MO$] \\
 \hline\\
  1 & $<$4 & $<$6 &        \\
  2 & $<$4 & $<$6 &        \\
  3 & ~~58 $^{\ast}$ &   82 & ~~$<$1.7 $^{\dagger}$ \\
  6 &      &      & $<$1.5 \\
  7 & $<$8 &$<$11 & $<$1.6 \\
 10 & $<$5 & $<$7 & $<$1.6 \\
 12 & $<$5 & $<$7 & $<$2.1 \\
 14 & $<$4 & $<$6 &        \\
 15 & $<$3 & $<$4 & $<$1.6 \\
 \hline\\
 \end{tabular*}

$^{\ast}$~
Only the narrow component of the H$\alpha$ emission is included.

$^{\dagger}$~
Maliolino et al.\ (1997) detected CO emission with a beam size of
$1'$.

\end{table*}

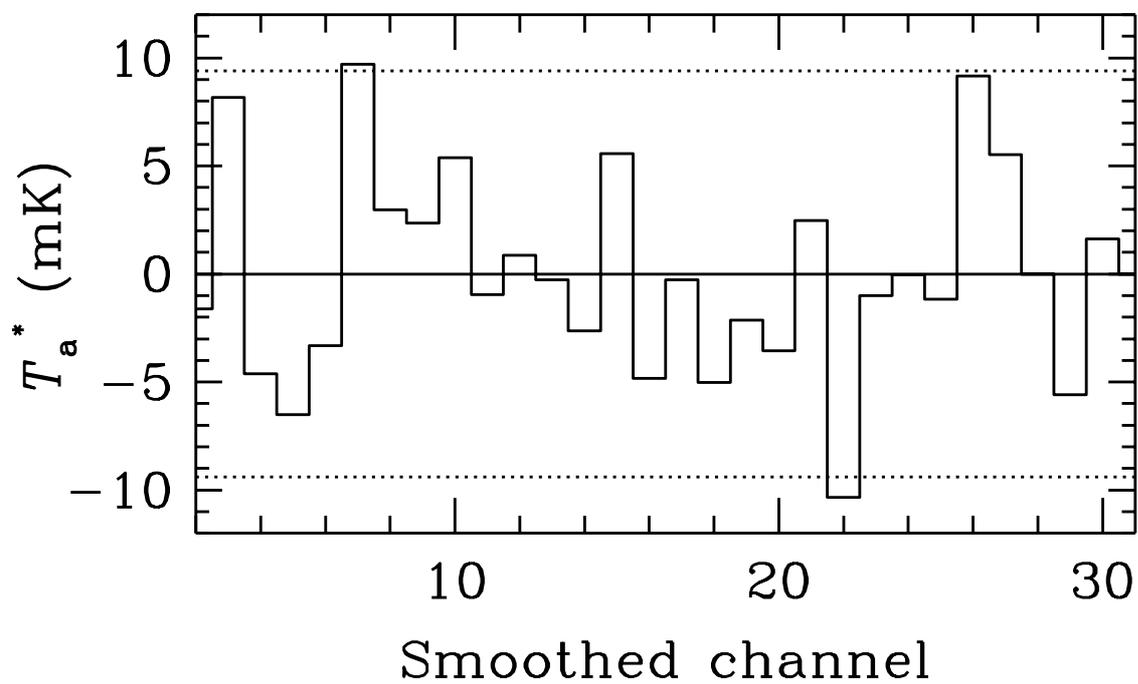
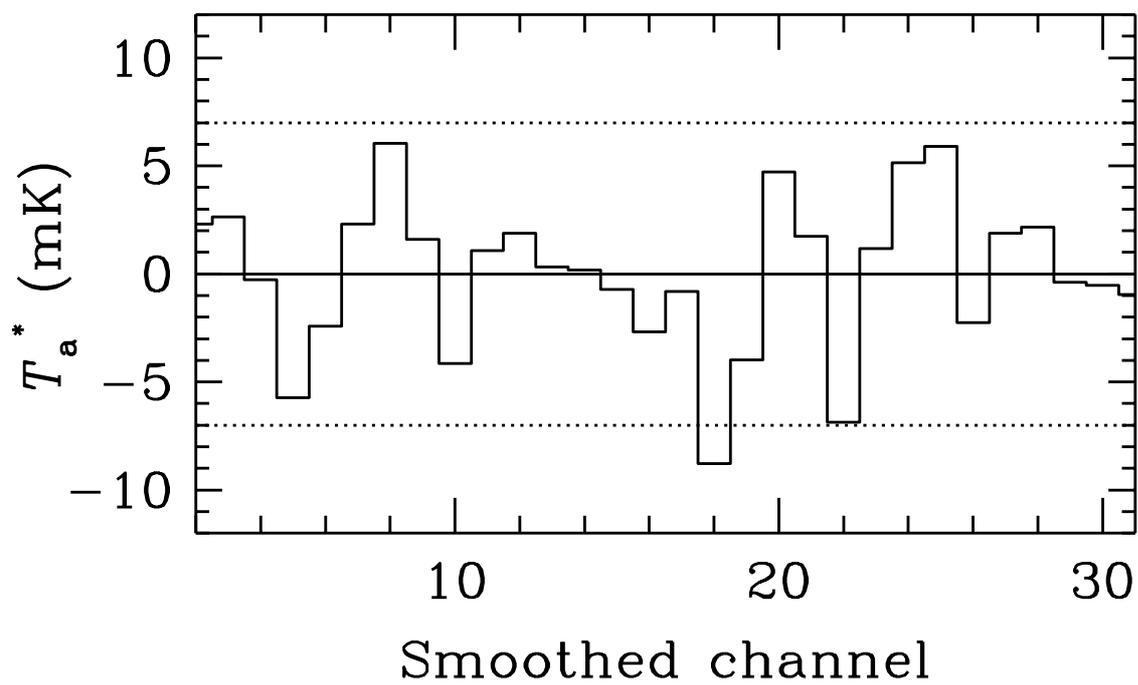
\end{document}